%% file: 0_main.tex
\documentclass{article}
\usepackage[T1]{fontenc} 
\usepackage[utf8]{inputenc} 
\usepackage{ismir,amsmath,cite,url,amssymb}
\usepackage{graphicx}
\usepackage{chngcntr}
\usepackage{subcaption}

\usepackage{array}
\newcommand{\PreserveBackslash}[1]{\let\temp=\\#1\let\\=\temp}
\newcolumntype{C}[1]{>{\PreserveBackslash\centering}p{#1}}
\newcolumntype{R}[1]{>{\PreserveBackslash\raggedleft}p{#1}}
\newcolumntype{L}[1]{>{\PreserveBackslash\raggedright}p{#1}}

\usepackage{lineno}

\title{Less is more: Faster and better music version identification with embedding distillation}

\multauthor
{Furkan Yesiler$^1$ \hspace{1cm} Joan Serr\`a$^2$ \hspace{1cm} Emilia G\'omez$^{3,1}$} {
  $^1$ Music Technology Group, Universitat Pompeu Fabra, Barcelona, Spain\\
$^2$ Dolby Laboratories, Barcelona, Spain\\
$^3$ European Commission, Joint Research Centre, Seville, Spain\\
{\tt\small furkan.yesiler@upf.edu}
}
\def\authorname{F. Yesiler, J. Serr\`a and E. G\'omez}

\begin{document}

\maketitle
\begin{abstract}
Version identification systems aim to detect different renditions of the same underlying musical composition (loosely called cover songs). By learning to encode entire recordings into plain vector embeddings, recent systems have made significant progress in bridging the gap between accuracy and scalability, which has been a key challenge for nearly two decades. In this work, we propose to further narrow this gap by employing a set of data distillation techniques that reduce the embedding dimensionality of a pre-trained state-of-the-art model. We compare a wide range of techniques and propose new ones, from classical dimensionality reduction to more sophisticated distillation schemes. With those, we obtain 99\% smaller embeddings that, moreover, yield up to a 3\% accuracy increase. Such small embeddings can have an important impact in retrieval time, up to the point of making a real-world system practical on a standalone laptop.
\end{abstract}
\input{1_introduction}
\input{2_related_work}

\input{3_methodology}
\input{4_results}
\input{5_conclusion}
\input{6_acknowledgment}

\bibliography{references.bib}
\input{7_appendix}


%
%
%
%
%

\end{document}

%% file: 1_introduction.tex
\section{Introduction}
The concept of music versions is as old as the concept of music itself. Before the existence of recorded music, listening to a piece mostly meant listening to a version of it. Nowadays, with the advancements in recording technologies, most music we listen to comes in recorded form. Nevertheless, musicians keep creating their own versions of existing songs for various reasons, including commercial ones (for example, to attract new audiences), political ones (to connect people or make a stance), and artistic ones (to re-imagine a song with a personal touch).

Version identification (VI) is the task of automatically detecting different renditions of the same underlying musical composition. VI systems are mainly focused on retrieval, aiming to find all renditions of a query song in a reference database. Although creating new versions is common practice, defining the characteristics that enable us to perceive the links connecting different renditions of the same piece is not a straightforward task~\cite{madoery2000, mosser2008}. Based on quantitative evidence, many successful systems exploit tonal and melodic descriptors that are invariant to the typical differences among versions, including the differences in timbre, tempo, structure, lyrics, and so on~\cite{serra2011, salamon2012, yesiler2020}. By further processing such descriptors, the ultimate goal is to obtain representations that allow inferring links among versions.

The accuracy-scalability trade-off stands as a key challenge in version retrieval. The early, alignment-based systems~\cite{serra2008, serra2009} incorporated musical know-how to capture the similarities among versions, resulting in strong performances. However, due to the scarcity of data, and their dependence on complex representations and computationally-intensive algorithms, they ended up in limited evaluation environments and, ultimately, not being suitable for industrial-size databases. With the release of the Million Song Dataset~\cite{millionsong}, researchers were further encouraged to address the scalability issue by exploring embedding-based systems that encode songs into more compact vectors. Although offering significant improvements for the scalability aspect, the performance of such systems failed to match their predecessors~\cite{yesiler2019}. Recent embedding-based systems that use deep learning techniques pave the way to encapsulate the similarities among versions in ways that are both efficient and accurate~\cite{doras2019, yu2019, yesiler2020, zalkow2020}.

The main use case of version retrieval in commercial settings is to detect copyright infringement cases in media streaming platforms and live performance venues or events. Such application scenarios require having fast and scalable solutions. For example, more than 500\,h of video content are uploaded to YouTube every minute\footnote{https://www.cnbc.com/2018/03/14/with-over-1-billion-users-heres-how-youtube-is-keeping-pace-with-change.html}, and handling the music licensing aspect of that requires having accurate and scalable systems that can identify the cases where a video includes a copyrighted piece of music. 

In this paper, we investigate a number of ways to improve the scalability of existing embedding-based VI systems that use neural networks as encoders. Specifically, our goal is to reduce the size of embedding vectors without compromising the accuracy of the systems. Since embeddings can be pre-computed, reducing their size is crucial to improve data storage and, more importantly, retrieval time. For this purpose, we consider three core state-of-the-art strategies, namely unsupervised dimensionality reduction, neural network pruning, and knowledge distillation. Apart from introducing a number of techniques from other fields to VI research, we also consider a novel knowledge distillation loss for metric learning, which aims to optimize a clustering evaluation metric. Moreover, inspired by transfer learning applications, we propose a technique called latent space reconfiguration, to show that learning a compact and efficient latent space is facilitated by using a pre-trained feature extractor due to its stronger priors, compared with using a randomly-initialized one. Our experiments suggest that the performance of a pre-trained network can be preserved, or even improved, while shrinking the embedding vectors down to less than 1\% of their original sizes. We evaluate our approach on a publicly-available test set, and share our code, instructions for using a newly-contributed training dataset and supplementary materials (SM) on Github\footnote{https://github.com/furkanyesiler/re-move}.


%% file: 2_related_work.tex
\section{Related Work}

\subsection{Version identification}
Like many other systems in music information retrieval (MIR), VI systems extract audio descriptors to obtain relevant information from signals, including mid-level ones, such as pitch class profiles (PCP)~\cite{ellis2007, serra2009, silva2016, tralie2017cover, chen2018} and predominant melody~\cite{foucard2010, salamon2012, doras2019}, or low-level ones, such as the constant-Q transform (CQT)~\cite{rafii2014, tsai2016, seetharaman2017}. To achieve invariance against the changes in musical characteristics, further processing steps have been proposed, including beat-synchronous features for handling tempo variations~\cite{ellis2007, bertin2012, tralie2017cover}, and the optimal transposition index for handling pitch transpositions~\cite{serra2008, serra2009, silva2016, tralie2017cover}. Many rule-based VI systems use alignment algorithms to then compare these representations, resulting in long retrieval times.

Embedding-based VI systems are designed to obtain compact representations that speed up the retrieval phase. Compact embeddings reduce the required storage and facilitate similarity estimation through the use of efficient nearest-neighbor libraries implementing common metrics like Euclidean or cosine distances. Early attempts of such systems use techniques like the 2D Fourier transform, principal component analysis, and linear discriminant analysis for encoding and dimensionality reduction operations~\cite{bertin2012, humphrey2013}.

Current deep learning-based systems learn non-linear transformations that map the feature representations into embedding vectors of various sizes, ranging from 300 to 16,000. Xu et al.~\cite{xu2018} and Yu et al.~\cite{yu2019} train their convolutional networks with a classification loss, but obtain the embedding vectors to use in the retrieval phase from the penultimate layer of their networks. Doras and Peeters~\cite{doras2019}, and Yesiler et al.~\cite{yesiler2020} formulate the network training as a metric learning setting, in which they use the triplet loss for optimizing distances among training samples.

\subsection{Metric learning}
This line of research is concerned with learning functions that produce low distance values between semantically similar data points, and high values otherwise. The early approaches include learning Mahalanobis distances~\cite{mahalanobis} with linear~\cite{goldberger2004, weinberger2006} or non-linear~\cite{kedem2012} transformations. Parametrizing such transformations with neural networks was pioneered by Chopra et al.~\cite{chopra2005} and Salakhutdinov and Hinton~\cite{salakhutdinov2007}, and the research domain combining these two concepts is often called deep metric learning. 


Deep metric learning methods offer new solutions regarding how to exploit the semantic relations among data, often by formulating or revising loss functions. The triplet loss~\cite{schroff2015}, similar to LMNN~\cite{weinberger2006}, manipulates the distances between genuine and impostor pairs with an energy-based approach. The ProxyNCA loss~\cite{proxynca}, similar to NCA~\cite{goldberger2004}, and NormalizedSoftmax loss~\cite{zhai2019}, similar to cross entropy loss, maximize the likelihoods of samples being close to particular class proxies. The group loss~\cite{elezi2019} incorporates the idea that similar elements should belong in the same class by using replicator dynamics~\cite{weibull1995}. Although there is a variety of deep metric learning losses, each one with distinctive advantages, the triplet loss variants remain the popular (and almost unique) choice in MIR research.

\subsection{Model reduction}
\subsubsection{Neural network pruning}
Pruning a large neural network can preserve the original performance while eliminating more than 90\% of its weights~\cite{lecun1990, hanson1989, han2015, frankle2018}. 
The main challenge is to identify the importance of connections and weights, and previous techniques explored the use of absolute weight magnitudes~\cite{hanson1989, han2015, frankle2018} and the Hessian of the loss function~\cite{lecun1990}. Pruning operations can be performed layer- or network-wise, in a one-shot or an iterative fashion, and combined with quantization or clustering. To the best of our knowledge, network pruning has not been considered for VI, nor further explored in MIR systems in general.



\subsubsection{Knowledge distillation}
Bucil\u{a} et al.~\cite{bucila2006}, and later Hinton et al.~\cite{hinton2015}, explored the idea where a small neural network (the student model) is trained with the guidance from a wide, deep, and better-performing network (the teacher model). In the metric learning context, some works explored this idea with a slightly changed formulation: Classical knowledge distillation methods use teacher networks to guide the students on individual examples, but metric learning methods exploit similarity relations among samples. For this, researchers proposed methods that match a number of properties between the embeddings obtained from the teacher and the student models, including the ranks of retrieved samples~\cite{chen2017}, distances between samples~\cite{park2019}, class likelihood distributions~\cite{han2019}, and absolute positions of embeddings in the latent space~\cite{yu2019metric}. With few exceptions~\cite{dali, wu2017}, distillation techniques are largely under-explored in MIR, and we believe that no attempt has been done within VI.

%% file: 3_methodology.tex
\begin{figure*}[tb!]
\includegraphics[width=\linewidth]{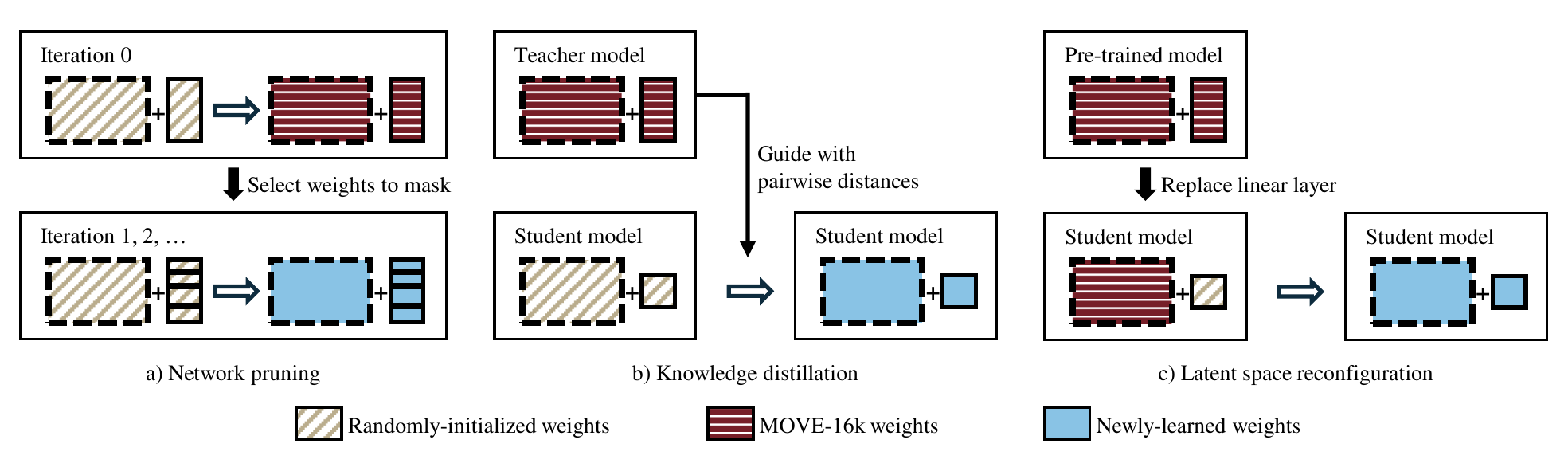}
\vspace{-8mm}
\caption{An overview of neural network-based embedding distillation methods. The hollow arrows denote training process, the boxes with dashed and with solid outlines denote feature extractors and fully-connected layers, respectively.}
\label{fig:methods}
\vspace{-3mm}
\end{figure*}

\section{Methodology}\label{sec:methods}
\subsection{Embedding distillation}
Our study focuses on a set of techniques for improving the scalability of existing VI systems in the retrieval phase by reducing the size of the embeddings, rather than building a novel network architecture.\footnote{To illustrate the benefits of using smaller embeddings, consider computing distances between a query and a reference database with 10\,M embeddings. This takes us (with a simple brute-force, double-loop Euclidean distance function) 0.32\,s using $d = 256$, but the elapsed time increases up to 4.75\,s for $d = 4k$, and to 18.43\,s for $d = 16k$ (the embedding size of MOVE~\cite{yesiler2020}). Although the absolute values are subject to change based on computational resources, for real-world applications on portable devices, such differences in magnitude for the retrieval time (from 0.32 to 18.43\,s) may drastically affect user experience and product appeal.}~We hypothesize that a high-capacity encoder is still needed to extract the essential information from complex and noisy signals such as current tonal representations. However, once a reliable encoder is obtained, it can be used for training a second model that outputs embeddings with a lower dimensionality, ideally without compromising accuracy. Due to the goal of having smaller embeddings that yield a similar performance, we call this set of methods embedding distillation techniques.


\subsection{Data}\label{sec:data}
Our models are developed with the Da-TACOS training set that we make available under Creative Commons BY-NC-SA~4.0 license together with this publication. It features a training partition of 83,904 songs in 14,499 cliques (or unique works), and a validation partition of 14,000 songs in 3,500 cliques. The annotations for the songs are obtained using the API of \url{secondhandsongs.com}. We share the crema-PCP features~\cite{mcfee2017structured} and the related metadata. 
Further detail on the contributed dataset is available in SM.

\subsection{Model architecture and training details}
Our methods require to start from a pre-trained and sufficiently reliable model. For this, we take advantage of the publicly-available MOVE model~\cite{yesiler2020}, together with its pre-trained weights. Nonetheless, we believe all methods introduced here can be applied to other embedding-based systems using neural networks (initial results are available in SM).

MOVE uses crema-PCP features $\textbf{X}\in [0,1]^{12\times T}$ as input, where $T$ is the number of frames, pre-computed with non-overlapping windows of 93\,ms. The model outputs embedding vectors $\textbf{v} = f(\textbf{X})\in\mathbb{R}^d$, where $d$ is the embedding size. The original work reports results for $d$ between 128 and 32\,k, and shows a clear accuracy drop for $d<2048$ (the final model employs a rather high dimensionality $d=16$\,k). In contrast, the dimensionalities we consider in this work are $d = \{128, 256, 512, 2048\}$.

To find a suitable learning rate and an optimizer for each experiment setting, we perform a grid search over both stochastic gradient descent and Ranger\footnote{https://github.com/lessw2020/Ranger-Deep-Learning-Optimizer} optimizers and initial learning rates in \{0.0001, 0.001, 0.01, 0.1\}, using our validation set. The full training lasts for 70~epochs, and we decrease the learning rate by a factor of 10 at epochs 50 and 60. We save the model weights that result in the best performance on the validation set. The remaining training details and design decisions follow the ones made by the MOVE authors~\cite{yesiler2020}. All neural network models are trained using PyTorch~\cite{pytorch}, and the hyper-parameter values used for each experiment can be found at our repository.

\subsection{Methods for embedding distillation}

\subsubsection{Classical unsupervised techniques}
Before going into complex solutions, we investigate the benefits of using classical dimensionality reduction techniques for embedding distillation. For this, we use principal component analysis (PCA), independent component analysis (ICA), and Gaussian random projection (GRP) techniques. Each model is fit using the training set embeddings obtained with MOVE-16k, and applied to the evaluation set embeddings. We use the implementations from the scikit-learn library~\cite{scikit-learn} and change only the number of target components.

\subsubsection{Pruning}
Based on the approach of Han et al.~\cite{han2015}, we study whether we can prune the dimensions of the latent space constructed by MOVE-16k in an iterative way. Although pruning the weights of all layers throughout the network is the most common practice, the underlying idea can be applied to only the final linear layer of the model in order to obtain embeddings with fewer dimensions. Denoting the weights of the linear layer of MOVE-16k as $\textbf{W}\in\mathbb{R}^{d\times f}$, where $d$ is the size of the embeddings and $f$ is the number of input connections to the linear layer, we compute the mean of the absolute values per row for $\textbf{W}$ and sort the rows based on these mean values. At the end of each iteration $m$ ($m\in\{0, 1, ... \}$), the weights of the top 50\% rows are restored to their initial values from Iteration 0 and retrained. The weights of the bottom 50\% rows are zeroed-out and not considered for the next iterations (\figref{fig:methods}(a)) .


\subsubsection{Knowledge distillation}
This set of experiments consider MOVE-16k as a teacher model, and our goal is to train from scratch a student model of the same size but with a lower embedding dimensionality. Our approach is formulated in a deep metric learning setting where the guidance of the teacher model is shaped by the distances among samples (\figref{fig:methods}(b)). In the experiments described next, the weights of the teacher model are frozen, and the weights of the student model are initialized randomly.

\vspace{1mm}

\noindent\textbf{Distance matching ---}
Perhaps the most intuitive way of guiding the student model is to match the distances obtained from the student with the ones from the teacher, allowing the two models to have different embedding sizes. In our implementation, we pass the samples in each mini-batch to both models, compute in-batch pairwise distances, and use the mean absolute error between the pairwise distance matrices from the teacher model and the student model to train the latter:
\begin{equation}\label{eq:distancematch}
L^{\text{DM}}_i = \sum_j \left| D(\textbf{v}_i^{\text{s}}, \textbf{v}_j^{\text{s}}) - D(\textbf{v}_i^{\text{t}}, \textbf{v}_j^{\text{t}}) \right| ,
\end{equation}
using
\begin{equation}\label{eq:dists}
D(\textbf{v}_i,\textbf{v}_j) = \frac{1}{d}~ \lVert \textbf{v}_i - \textbf{v}_j\rVert^2 ,
\end{equation}
where $\lVert~\rVert$ represents the Euclidean norm, and $\textbf{v}_i^{\text{s}}$ and $\textbf{v}_i^{\text{t}}$ the embeddings of song $i$ obtained with the student and teacher models, respectively.

\vspace{1mm}

\noindent\textbf{Cluster matching ---}
Our second knowledge distillation scheme aims to obtain a student model that constructs clusters with both low intra-class and high inter-class distances. Assuming the teacher model holds this desired property, we take advantage of this information to guide the student model. To the best of our knowledge, this distillation criterion has not been explored in previous deep metric learning research.

Our criterion exploits internal cluster evaluation metrics~\cite{liu2010}. In the experiments reported here, we use the Davies-Bouldin (DB) index~\cite{dbindex}, but other cluster evaluation metrics can be used:
\begin{equation}\label{eq:dbindex}
L^{\text{DB}}_i = \max_{j \neq i}\left(\frac{\sigma_i + \sigma_j}{D(c_i, c_j)}\right),
\end{equation}
where $\sigma_i$ denotes the average intra-class distance, computed with a suitable distance measure $D$, and $c_i$ denotes the centroid for class $i$. The DB index yields low values for configurations that have low intra-class and high inter-class distances. 

In our implementation, we pre-compute the class centroids using the MOVE-16k embeddings from the entire training set. To match the dimensions of the centroids with the student model embeddings, we train a linear projection simultaneously with the student model. The intra-class and inter-centroid distances are computed using only the samples present in the mini-batch and their respective centroids. After computing DB scores for each class in the mini-batch, we average them to obtain the final loss value. 


\subsubsection{Latent space reconfiguration}
Transfer learning applications take advantage of the strong priors learned by the feature extractor part of successful, high-capacity models that are trained on large datasets. Inspired by this idea, we hypothesize that, by using the feature extractor of a pre-trained model, we can obtain a better-structured and lower-dimensional latent space that cannot be obtained by training a randomly-initialized model from scratch. 

To test this idea, we use the pre-trained convolutional layers of MOVE-16k as the feature extractor, remove the final linear layer, and learn a new latent space with a randomly-initialized linear layer using a metric learning loss function (\figref{fig:methods}(c)). Note that the original MOVE-16k model is trained with a triplet loss, meaning that it learned a distance metric parametrized by a neural network. Our approach uses the non-linear part of that metric, and `reconfigures' the latent space and the distance metric by optimizing a second loss function (hence the name latent space reconfiguration). Our motivation is based on two assumptions: (1) training losses play an important role in shaping the latent space where the embeddings lie, and (2) embeddings with lower dimensionalities may be sufficient to successfully represent semantically meaningful information, as long as the dimensions are effectively utilized. 
Note that although this technique follows the same procedure as transfer learning, the latter requires, by definition, distinct source and target tasks (or datasets), which is not the case for the proposed technique. Focusing on metric learning schemes, the term latent space reconfiguration denotes the process of starting with an already learned distance metric and modifying it to represent the semantic relations in a more compact embedding space.

In our experiments, we consider 4~loss functions which are explained below. The weights of the feature extractor are frozen during the first epoch and updated with a lower learning rate during the rest of the training. Batch normalization is applied after the linear layer as in MOVE-16k. Apart from using the loss functions below for latent space reconfiguration, we also use them individually and train models from scratch with the same settings to set baseline models. 

\vspace{1mm}

\noindent\textbf{Triplet loss ---} 
We follow the triplet loss formulation used by Yesiler et al.~\cite{yesiler2020}. Distances among vectors are computed using $D$ as specified in \eqnref{eq:dists}:
\begin{equation}\label{eq:triplet}
L^{\text{T}}_i = \max \left( D\left(\textbf{v}_i,\textbf{v}_+\right) - D\left(\textbf{v}_i,\textbf{v}_-\right) + m, 0\right),
\end{equation}
where $\textbf{v}_i$ corresponds to the anchor, $\textbf{v}_+$ to the positive sample, $\textbf{v}_-$ to the negative sample, and $m=1$ is a margin hyper-parameter. For selecting which triplets to use, we follow the hard-positive, hard-negative mining strategy used by the authors. 

\vspace{1mm}

\noindent\textbf{ProxyNCA loss ---}
Our implementation of ProxyNCA loss~\cite{proxynca} also uses the normalized squared Euclidean distance metric from \eqnref{eq:dists}. Every class in our training set is represented with one proxy vector that is initialized randomly and trained simultaneously with the model parameters. In mathematical notation, the ProxyNCA loss can be expressed as:
\begin{equation}\label{eq:proxynca}
L^{\text{P}}_i = - \log\left(\frac{\exp(-D(\textbf{v}_i, y))}{\sum_{z \in Z} \exp (-D(\textbf{v}_i, z))}\right),
\end{equation}
where $y\in\mathbb{R}^d$ denotes the proxy vector for the class of $\textbf{v}_i$ and $Z$ denotes the set of proxies for all classes different than the one of $\textbf{v}_i$. 

\vspace{1mm}

\noindent\textbf{NormalizedSoftmax loss ---}
As proposed in~\cite{zhai2019}, we implement this function using the cosine distance. We randomly initialize one proxy per class and update their parameters at each training step. We use
\begin{equation}\label{eq:normsoftmax}
L^{\text{N}}_i = - \log\left(\frac{\exp(\langle \textbf{v}_i, y \rangle / \tau)}{\sum_{z \in Z} \exp(\langle \textbf{v}_i, z \rangle / \tau)}\right),
\end{equation}
where $\langle~\rangle$ denotes cosine similarity, $y\in\mathbb{R}^d$ the proxy for the positive class, $Z$ the set of proxies for all classes, and $\tau=0.05$ the temperature parameter.

\vspace{1mm}

\noindent\textbf{Group loss ---}
Following the approach of~\cite{elezi2019}, we use Pearson’s correlation coefficient as the similarity metric and replace the negative values with 0. We perform three iterations for refining the class probabilities and, unlike the original implementation, we select one anchor per class in the mini-batch. The main loss is regular cross-entropy: 
\begin{equation}\label{eq:group}
L^{\text{G}}_i = - \log\left(\frac{\exp(l^{c}_i)}{\sum_{t \in C} \exp(l^{t}_i)}\right),
\end{equation}
where $l^{c}_i$ denotes the logit of sample $i$ with respect to its positive class $c$, and $C$ denotes the set of all classes in the training set. However, in group loss, logits are updated with replicator dynamics using pairwise similarities~\cite{elezi2019}. 



%% file: 4_results.tex
\section{Results}
\subsection{Evaluation methodology}
For development, we use the newly available dataset mentioned in Section~\ref{sec:data} and detailed in SM. Results are then evaluated on Da-TACOS benchmark subset~\cite{yesiler2019}, which contains a non-intersecting set of cliques with respect to our training and validation data. Da-TACOS contains 1,000 cliques with 13 songs each and 2,000 noise songs that do not belong to any other clique and are not queried. Following common practice, we report the performance of our models using mean average precision (MAP) and mean rank of the first relevant item (MR1) metrics.



\subsection{Embedding distillation experiments}\label{sec:embdist_res}
Table~\ref{tab:methods} presents the exhaustive list of results for the methods described in Section~\ref{sec:methods}. The baseline results (top block) show that, when training from scratch, changing the loss function of a network causes significant accuracy differences. It should be noted that all alternative losses we consider achieve state-of-the-art performances in computer vision datasets. Nevertheless, our results suggest that they may not generalize across other types of data or tasks, or that they may be oversensitive to hyper-parameters or specific architectural decisions.

\begin{table}[tb!]
\begin{center}
\resizebox{\columnwidth}{!}{
\begin{tabular}{l c c c c}
\hline\hline
Method & \multicolumn{4}{c}{$d$}\\
 & 128 & 256 &  512 & 2048\\
\hline\hline
\multicolumn{5}{l}{\textit{Baselines (no reduction, training from scratch)}}\\
Triplet  & 0.459 & 0.469 &	0.478 &	0.487\\

ProxyNCA  & 0.168 &0.185 &	0.212 &	0.206\\

NormalizedSoftmax  & 0.445 & 0.470	& 0.475 &	0.422\\

Group  & 0.265 &0.271	&0.269 &	0.271\\

\hline
\multicolumn{5}{l}{\textit{Unsupervised}}\\
PCA & 0.494 & \textbf{0.507} & \textbf{0.507}	& \textbf{0.507} \\
ICA & 0.456 & 0.425 & n/a & n/a \\
GRP & 0.429 & 0.465 & 0.485	& \textbf{0.502} \\

\hline
\multicolumn{5}{l}{\textit{Knowledge distillation}}\\

Distance matching + Triplet & 0.492 & \textbf{0.499} & \textbf{0.503}	& \textbf{0.500} \\

Cluster matching + Triplet  & 0.424 & 0.471 &	0.465 &	0.455\\
\hline

\multicolumn{5}{l}{\textit{Latent space reconfiguration}}\\
Triplet  & 0.485 & 0.491 &	0.494 &	\textbf{0.506}\\

ProxyNCA  & 0.424 &0.465 &	0.485 & \textbf{0.502}\\

NormalizedSoftmax  & \textbf{0.513} & \textbf{0.524} &	\textbf{0.525} &	\textbf{0.525}\\

Group  & 0.465 & 0.483	& \textbf{0.495} &	\textbf{0.511}\\

\hline\hline
\end{tabular}
}
\caption{MAP for different embedding sizes $d$ when training from scratch (top) and when using pre-trained models and embedding distillation (middle-bottom). MAPs for the original MOVE-4k and MOVE-16k baselines are 0.495 and 0.507, respectively (values equal to or above MOVE-4k are highlighted in bold).}\label{tab:methods}
\end{center}
\end{table}

For unsupervised dimensionality reduction (second block of Table~\ref{tab:methods}), we find that PCA successfully projects the information contained in MOVE-16k embeddings, even when using 256~dimensions. This suggests that, although achieving state-of-the-art performance, MOVE-16k embeddings contain redundant information that can be drastically compressed. GRP reaches a similar performance as PCA with $d=2048$, but the resulting performance decreases when using lower-dimensional embeddings.

The initial experiments on pruning reached the same performance as MOVE-16k after one iteration, that is, after reducing the dimensionality by 50\%. However, further pruning iterations drastically decreased MAP, up to the point of yielding non-useful embeddings. Therefore, we decided to stop iterating and not to report the corresponding results.

Among the considered knowledge distillation techniques (third block, Table~\ref{tab:methods}), the additional distance matching loss clearly increases the model performance compared with the case where only the triplet loss is optimized. However, the same advantage is not observed with cluster matching using DB loss. We hypothesize that this may be related to training an extra linear projection for compressing the centroid embeddings to match the size of the embeddings obtained with the student model.

Latent space reconfiguration results seem to justify our hypothesis regarding the use of strong priors of a pre-trained feature extractor (last block, Table~\ref{tab:methods}). All considered alternatives outperform their baseline counterparts. Moreover, we find that using probabilistic losses such as NormalizedSoftmax and Group for latent space reconfiguration even outperforms the original model while reducing the embedding size by a large margin ($128/16000=0.8\%$). Notice that, in addition to these advantages, latent space reconfiguration does not suffer from the setbacks of network pruning and knowledge distillation methods, namely training a model for multiple iterations and using two models simultaneously during training, respectively.

\begin{table}[tb!]
\setlength\tabcolsep{12pt}
\begin{center}
\begin{tabular}{l r c c}
\hline\hline
 & \multicolumn{1}{c}{$d$} & MAP &  MR1\\
\hline\hline


2DFTM~\cite{bertin2012}  & 50 & 0.275  & 155 \\

Dmax~\cite{chen2018}  & $5.5\,\text{k}$ & 0.322 & 132 \\

SiMPle~\cite{silva2016}  & $2.2\,\text{k}$ & 0.332 &  142\\

Qmax~\cite{serra2009}  & $5.5\,\text{k}$ & 0.365   & 113  \\

Qmax*~\cite{serra2009unsupervised}  & $5.5\,\text{k}$ & 0.373 & 104 \\

EarlyFusion~\cite{tralie2017cover}  & $8.5\,\text{k}$ & 0.426  &   116 \\

LateFusion~\cite{chen2018}  & $5.5\,\text{k}$ & 0.454  &   177 \\

MOVE~\cite{yesiler2020} & $4\,\text{k}$ & 0.495 & 42 \\ 

MOVE~\cite{yesiler2020} & $16\,\text{k}$ & 0.507 & 40 \\ 


\textbf{Re-MOVE} & \textbf{256} & \textbf{0.524} & \textbf{38} \\ 

\hline\hline
\end{tabular}
\caption{Comparison with existing VI systems using Da-TACOS (taken from~\cite{yesiler2019}). When not explicit, embedding sizes $d$ are estimated for a song duration of 3.5\,min (see text). Results for the proposed methodology are highlighted in bold.}\label{tab:sota}
\end{center}
\end{table}

\subsection{Comparison with the state of the art}
Lastly, Table~\ref{tab:sota} compares our best result with state-of-the-art methods. The second column, $d$, shows the size of the smallest representation (per song) required for each method to estimate pairwise similarities (equivalent to the embedding dimensionality). As the results for the first 7~methods are computed with the publicly-available acoss library~\cite{yesiler2019}, we use those implementations for estimating the embedding sizes\footnote{For 2DFTM, the acoss library uses a 450-dimensional embedding while the authors apply PCA to reduce the dimensionality to 50.}. As the sizes of some representations depend on the song duration (SiMPle, Qmax, Dmax, LateFusion) or tempo (EarlyFusion), we use 3.5\,min and 102\,bpm estimates, which correspond to average song duration and bpm of the songs in Da-TACOS, respectively.

\begin{figure}[tb!]
\includegraphics[width=0.97\linewidth]{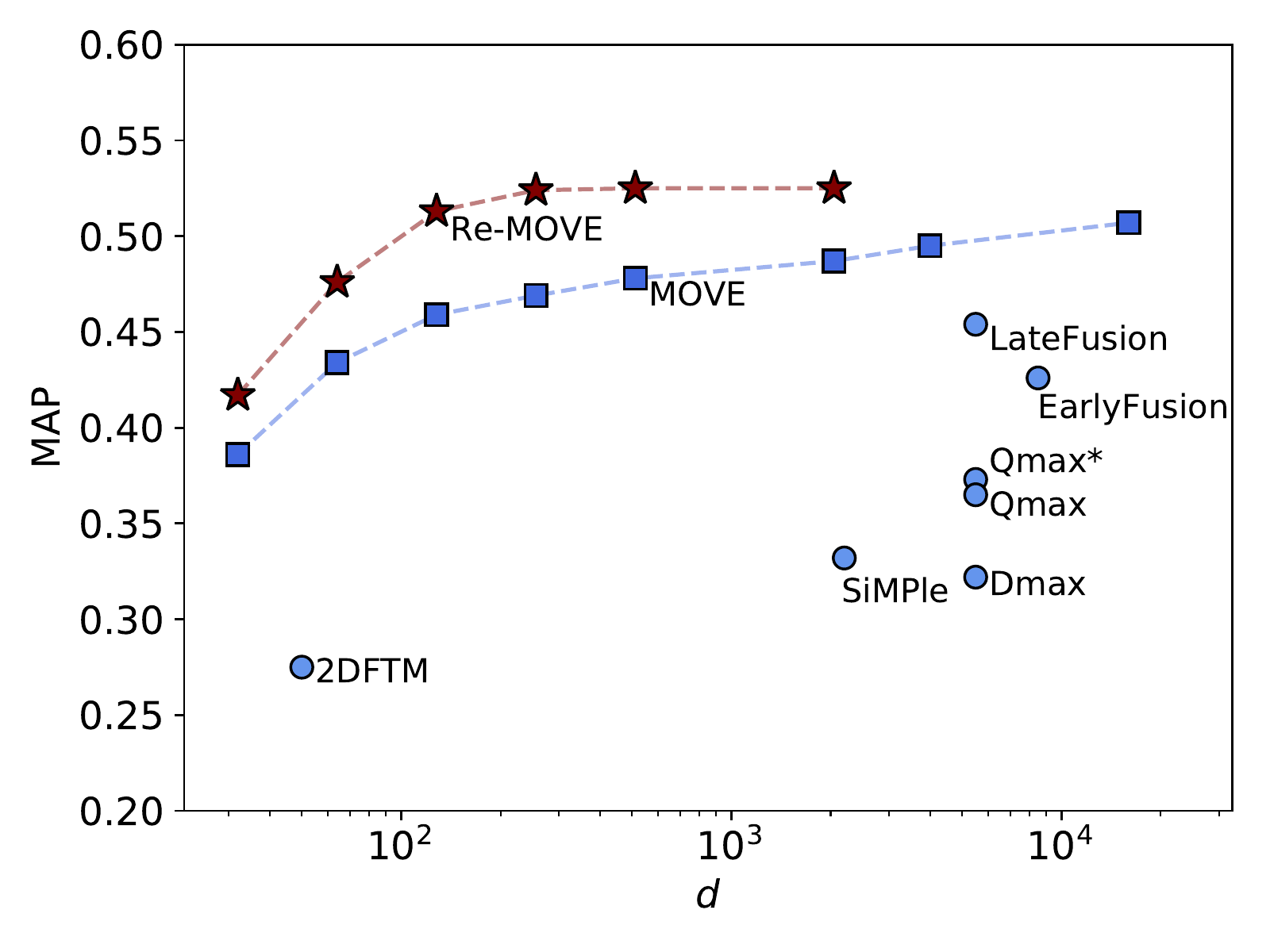}
\vspace{-5mm}
\caption{MAP with respect to embedding dimensionality $d$ for Re-MOVE (red stars), MOVE (blue squares), and other existing approaches (blue circles). Notice the logarithmic axis.}
\label{fig:sota}
\end{figure}

Re-MOVE, which stands for `reduced MOVE', denotes the model trained with latent space reconfiguration using NormalizedSoftmax. With $d=256$, it demonstrates relative performance increases of 3\%, 6\%, and 15\% when compared with MOVE-16k, MOVE-4k, and LateFusion systems, respectively (Table~\ref{tab:sota}). We also find that Re-MOVE improves over MOVE for a wide range of dimensionalities $d\in[32,2048]$ (Figure~\ref{fig:sota}). Along with its state-of-the-art performance, Re-MOVE provides a crucial advantage in terms of scalability, which positions it as the most viable system from a practical point of view.

%% file: 5_conclusion.tex
\section{Conclusion}
In this work, we have studied a set of techniques for reducing the embedding sizes of existing VI systems, which we consider under the name embedding distillation. We have claimed that by using a pre-trained and high-capacity network, we can train a second network that yields smaller embedding vectors without a decrease in performance. To investigate this idea, we have studied a wide range of techniques, including classical dimensionality reduction, neural network pruning, and knowledge distillation methods. Moreover, we have introduced latent space reconfiguration, which is a technique that builds upon the non-linear part of a distance metric learned by a pre-trained network to construct a compact latent space with fewer dimensions. Our results show that it is possible to reduce the embedding dimensionality of a model while maintaining, or even surpassing, its performance.

As future work, we plan to investigate further techniques for compressing entire networks rather than just embedding vectors. We emphasized the importance of having smaller embeddings for real-world applications, and we plan to demonstrate it further in carefully-designed version retrieval scenarios that mimic real-world use cases. Lastly, we believe that optimizing the existing methods to make them applicable in industrial scenarios is a valuable research direction, and we hope to facilitate bridging the gap between academy and industry in MIR research.

%% file: 6_acknowledgment.tex
\clearpage
\section{Acknowledgments}
This work is supported by the MIP-Frontiers project, the European Union's Horizon 2020 research and innovation programme under the Marie Skłodowska-Curie grant agreement No.~765068, and by TROMPA, the Horizon 2020 project 770376-2.

%% file: 7_appendix.tex
\clearpage\onecolumn
\appendix
\renewcommand{\thetable}{S\arabic{table}}  
\renewcommand{\thefigure}{S\arabic{figure}}
\setcounter{figure}{0}
\setcounter{table}{0}

\begin{center}
\textbf{SUPPLEMENTARY MATERIALS}
\end{center}

\section{D\lowercase{a}-TACOS Training set}

As mentioned in Section~\ref{sec:data}, we publicly share the Da-TACOS training set, which includes pre-computed crema-PCP features and the related metadata for a total of 97,904 songs in 17,999 unique works or cliques. The training and validation partitions include randomly-selected, disjoint sets of 14,499 and 3,500 cliques, respectively. The average duration of the songs is 218\,s, with a standard deviation of 69\,s. All audio files used for computing the features are encoded in MP3 format with varying bit rates, and their sample rate is 44.1 kHz. The metadata and the annotations are obtained using the API of \url{secondhandsongs.com}, and they are shared under the Creative Commons BY-NC~3.0 license. For each song, the shared metadata includes the song title and artist, the work title and artist, release year, SecondHandSongs.com performance and work IDs, and whether the song is instrumental or not.

Together with SHS-100K~\cite{xu2018}, this is one of the largest datasets available for training VI systems. An important advantage of the current dataset is that it contains a completely disjoint set of cliques with respect to the publicly-available evaluation set Da-TACOS~\cite{yesiler2019}. Therefore, researchers can use it as their new training and validation datasets and still report their test results using Da-TACOS benchmark subset.

\begin{figure}[ht!]
\begin{subfigure}{.495\linewidth}
    \centering
    \includegraphics[width=\linewidth]{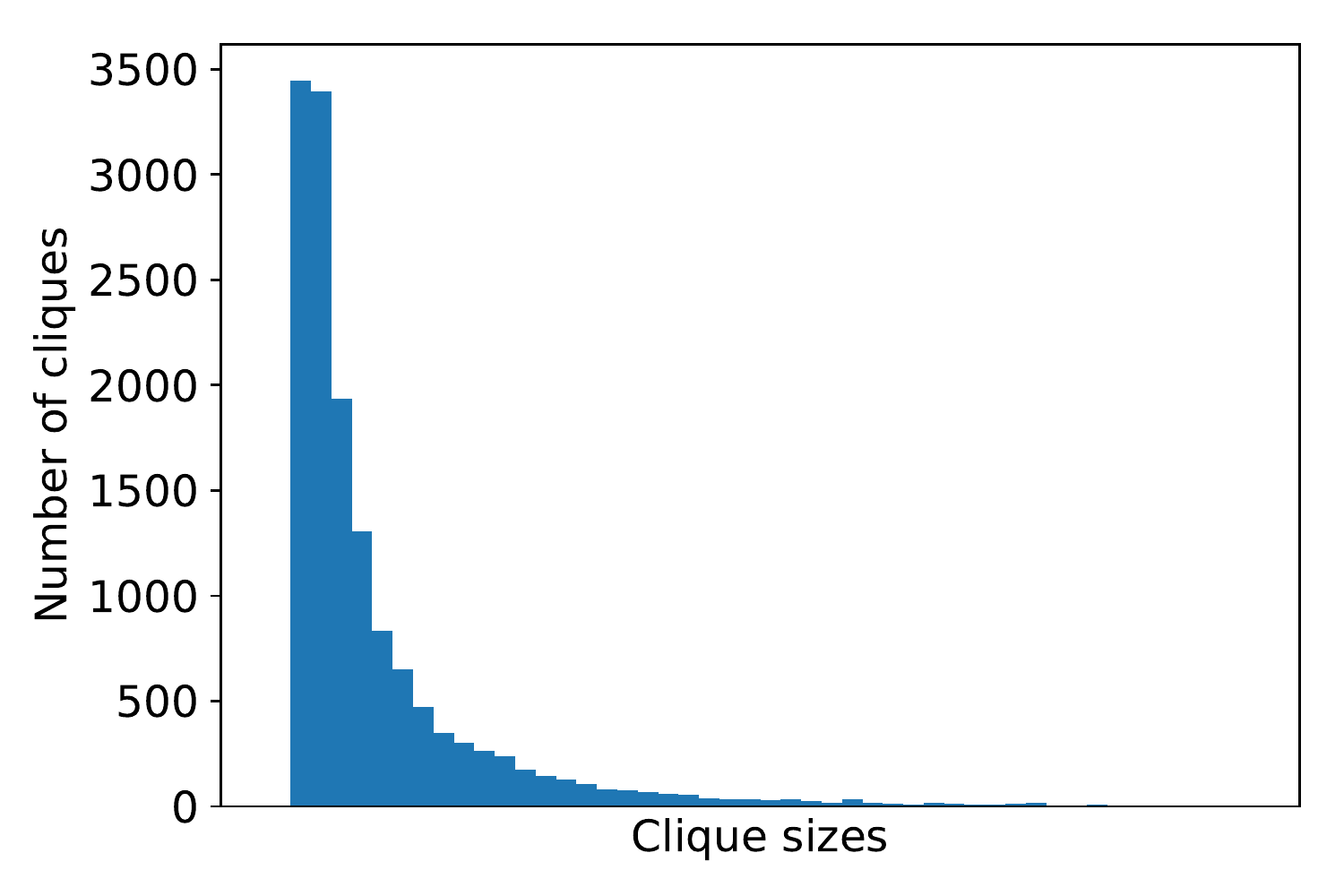}
\end{subfigure}
\begin{subfigure}{.495\linewidth}
    \centering
    \includegraphics[width=\linewidth]{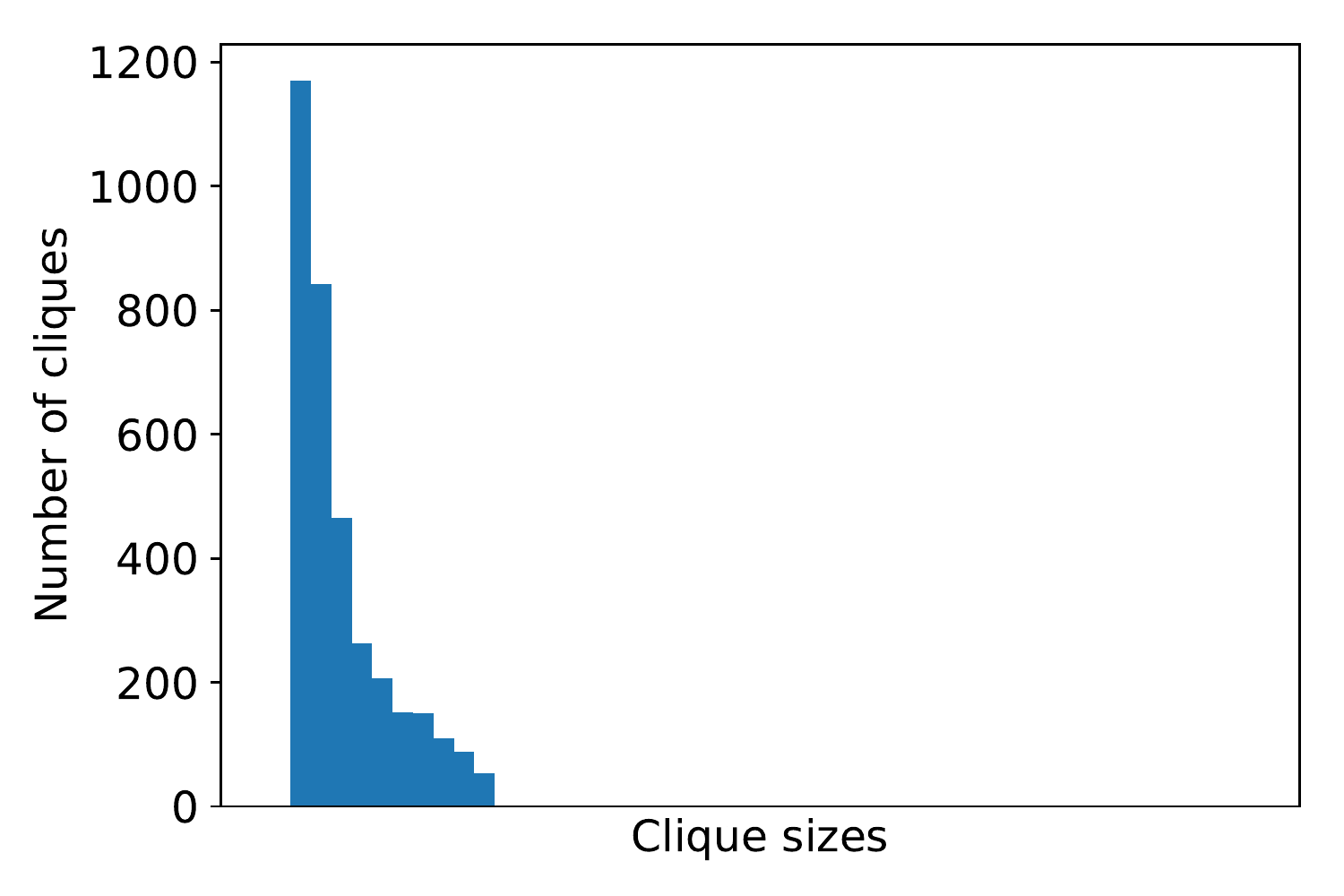}
\end{subfigure}
\vspace{-2mm}
\caption{Distribution of the number of songs per clique for the training (left) and the validation (right) sets.}
\label{fig:dists}
\end{figure}

\figref{fig:dists} shows the distributions of songs per clique for the training (83.9\,k songs, left) and the validation partitions (14.0\,k songs, right). The number of songs per clique ranges from 2 to 109 in the training set, and from 2 to 11 in the validation set. Our intention was to mimic real-world data where it is more likely to have more unique works with less number of songs, rather than having a balanced dataset in terms of the number of songs per clique. 

\section{Generalizability of latent space reconfiguration}
The experimental results reported in the main paper show that the latent space reconfiguration technique takes advantage of using a pre-trained feature extractor to reach high accuracies with fewer dimensions compared with using a randomly-initialized model. We claim that due to its model-agnostic nature, it can be used with any neural network-based VI system to demonstrate a similar performance gain. 

To test our claim, we implement the encoder proposed by Doras and Peeters~\cite{doras2019}, and investigate whether we can obtain similar results compared to the ones in Section~\ref{sec:embdist_res}. For this, we extracted the dominant melody representations used in~\cite{doras2019} for both our training data and Da-TACOS. To start with, we need a reliable base model (as MOVE-16k) for using its feature extractor. We train the network proposed in~\cite{doras2019} with $d = 512$ and call it F0-512 for readability. To show the effect of using latent space reconfiguration, we later train the same network using $d = \{64, 128\}$ from scratch. Lastly, we use the feature extractor of F0-512, remove the linear layer and learn a new latent space with randomly-initalized linear layer using $d = \{64, 128\}$. Compared to the experiments in the main paper, F0-512 is the counterpart of MOVE-16k, and the networks with $d = \{64, 128\}$ correspond to the newly-trained MOVE models with $d < 2048$. For this set of experiments, we consider only the triplet loss in order to isolate the effect of using latent space reconfiguration from other factors such as the performance of different loss functions.

Instead of replicating their entire training setting, we use the training strategies explained for our other experiments, with a few exceptions: (1) we use Adam optimizer instead of SGD or Ranger, (2) the mini-batch size is 100, (3) we do not apply data augmentation to the samples, (4) the latent space reconfiguration experiments use an initial learning rate 0.1 while the others use 0.0001 (using 0.1 for training from scratch resulted in collapsed models). 

\begin{table}[h!]
\setlength\tabcolsep{9pt}
\begin{center}
\resizebox{0.4\columnwidth}{!}{
\begin{tabular}{L{3cm} C{1cm} C{1cm}}
\hline\hline
Method & \multicolumn{2}{c}{$d$}\\
 & 64 & 128 \\
\hline\hline
\multicolumn{3}{l}{\textit{Baselines (training from scratch)}}\\
Triplet  & 0.283 & 0.279 \\

\hline

\multicolumn{3}{l}{\textit{Latent space reconfiguration}}\\
Triplet  & 0.298 & 0.295 \\

\hline\hline
\end{tabular}
}
\caption{MAP for different embedding sizes $d$ when training from scratch (top) and when using latent space reconfiguration (middle-bottom). MAP for F0-512 is 0.277. }\label{tab:f0_lsr}
\end{center}
\vspace{-1mm}
\end{table}

\tabref{tab:f0_lsr} presents our findings. Firstly, the baseline results (top) appear counter-intuitive as using lower embedding sizes results in higher performances. We believe that this may be related to not optimizing the training strategies and hyper-parameters for this specific architecture using this specific input representation. Secondly, the latent space reconfiguration results (bottom) seem consistently superior compared with the baseline results. The relative increases with respect to their baseline counterparts are 5.3\% and 5.7\% for $d=64$ and $d=128$, respectively. These results suggest that the performance increase achieved by using latent space reconfiguration is not limited to the network architecture or the input representation used for the experiments reported in the main paper. However, a more comprehensive analysis using several other network architectures and datasets is still needed to conclude that latent space reconfiguration technique can be applied to any embedding-based VI system.